\documentstyle[multicol,aps,epsf]{revtex}

\begin{document}
\title{Clogging Time of a Filter}
\author{S.~Redner and Somalee~Datta$^*$}

\address{Center for BioDynamics, Center for Polymer Studies, and Department of Physics, Boston
University, Boston, MA, 02215}
\maketitle
\begin{abstract}
  
  We study the time until a filter becomes clogged due to the trapping of
  suspended particles as they pass through a porous medium.  This trapping
  progressively impedes and eventually stops the flow of the carrier fluid.
  We develop a simple description for the pore geometry and the motion of the
  suspended particles which, together with extreme-value statistics, predicts
  that the distribution of times until a filter clogs has a power-law
  long-time tail, with an infinite mean clogging time.  These results and its
  consequences are in accord with simulations on a square lattice porous
  network.

\end{abstract}
\begin{multicols}{2}
  
  In this letter, we investigate the time required for a filter to clog.  In
  a typical filtration process, a dirty fluid is ``cleaned'' by passing it
  through a porous medium to remove the suspended particles.  The medium
  enhances filtering efficiency by increasing both the available filter
  surface area for trapping suspended particles, as well as the exposure time
  of the suspension to the active surfaces.  Such a mechanism is the basis of
  water purification, air filtration, and many other separation
  processes\cite{rev,rev2}.  As suspended particles become trapped, the fluid
  permeability of the medium gradually decreases, and eventually the filter
  becomes clogged.  Determining the time dependence of this clogging is basic
  to predicting when a filter is no longer useful, either because of reduced
  throughput or reduced filtration efficiency, and should be discarded.
  
\begin{figure}
  \narrowtext \epsfxsize=2.6in\epsfysize=2.3in \hskip
  0.2in\epsfbox{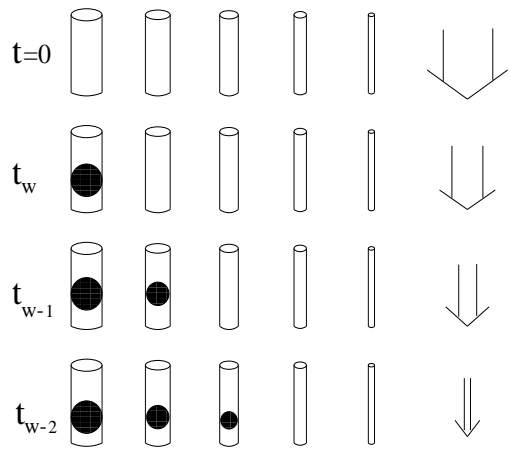} \vskip 0.15in
\caption{Illustration of the clogging evolution in a parallel array of 
  $w$ bonds of distributed radii.  Bonds are blocked in decreasing size
  order.  The overall flow decreases more slowly in the latter stages
  (vertical arrow).  The total flow is inversely related to the time until
  each blockage event, $t_k$, with $k=w,w-1,\ldots,$.
\label{cartoon}}
\end{figure}

  We develop a minimalist model which provides an intuitive understanding for
  the clogging of a filter.  Our analytical results are based on representing
  the clogging of a porous medium by the clogging of a single parallel array
  of pores which are blocked in decreasing size order (Fig.~\ref{cartoon}).
  This geometrical reduction is based on previous filtration studies which
  showed that, when particles and pores are of comparable size, clogging
  preferentially occurs in the upstream end of the network\cite{sch,dr}.  A
  single parallel array of pores represents the ultimate limit of this
  gradient-controlled process.
  
  The approximation of size-ordered blocking is based on the commonly-used
  picture that a particle has a probability proportional to the relative flux
  to enter a given unblocked pore from among the outgoing pores at a junction
  \cite{rev2,sch,sahimi}.  For Poiseuille flow, this flux is proportional to
  $r^4$, where $r$ is the pore radius.  Size-ordered blocking arises if the
  exponent 4 in this flow-induced pore entrance probability is replaced by
  $\infty$.  For a broad distribution of pore radii, this is not a drastic
  approximation.  This ordering also provides a direct correspondence between
  the radius of the currently-blocked pore and the time until it is blocked.
  We then use extreme value statistics\cite{extreme} to compute the radius
  distribution of the last few unblocked pores, together with the connection
  between the radius of the currently blocked pore and the network
  permeability, to determine the distribution of filter clogging times.  The
  predictions of this simple modelling are in good agreement with Monte Carlo
  simulations of clogging on a square lattice porous network.
    
In our simulations of clogging, we consider a lattice porous medium whose
bonds represent pores.  We assume Poiseuille flow, in which the fluid flux
through a bond of radius $r_i$ is proportional to $-r_i^4\,\nabla p$, where
$\nabla p$ is the local pressure gradient when a fixed overall pressure drop
is imposed.  Dynamically neutral suspended particles traverse the medium in
accordance with the local flow and the flow-induced entrance probability at
each junction.  A crucial feature is that particles are injected at a {\em
  finite} rate which is proportional to the overall fluid flux, corresponding
to a fixed non-zero density of suspended particles in the fluid\cite{sahimi}.
This is in distinction to many previous studies of filtration, where
particles were injected singly and tracked until trapping
occurred\cite{rev2,sch,dr}; this corresponds to an arbitrarily dilute
suspension.  While useful for understanding the percolation process induced
by clogging\cite{sch,dr,sahimi,perc}, the time evolution of filtration
necessitates the consideration of a finite-density suspension.
  
For the trapping mechanism we adopt size exclusion, in which a particle of
radius $r_{\rm particle}$ is trapped within the first bond encountered with
$r_{\rm bond}<r_{\rm particle}$\cite{sahimi}.  We assume that a trapped
particle blocks a pore completely and permanently.  The difference between
partial and total blockage in a single trapping event appears to be
immaterial for long-time properties\cite{dr}.  The overall trapping rate is
then controlled by the relative sizes of particles and bonds.  For simplicity
and because it occurs in many types of porous media\cite{rmp}, we consider
the Hertz distribution of pore and particle radii, respectively,
\begin{equation}
\label{radii}
b(r) =2\alpha r e^{-\alpha r^2}, \qquad {\rm and}\qquad p(r) =2 r e^{- r^2},
\end{equation}
with the ratio between the average bond and particle radii
$s=1/\sqrt{\alpha}$ a basic parameter which determines the nature of the
clogging process.

First consider the case where pores are larger than particles ($s\agt 1$).
Many particles must be injected before a sufficiently large particle arises
which can block the largest pore.  Since the permeability of the system
remains nearly constant when only a few pores are blocked, the time between
successive particle injection events during this initial stage is nearly
constant.  Once the largest pores are blocked, it takes many fewer particles
to block the remaining smaller pores and clogging proceeds more quickly.
Thus the clogging time is dominated by the initial blockage events.

To estimate the clogging time in this large-pore limit, let us find the
number of injection events before encountering a particle large enough to
block the largest bond.  The radius of this bond is given by the
criterion\cite{extreme}
\begin{equation}
\label{rmax}
\int_{r_{\rm max}}^\infty 2\alpha r e^{-\alpha r^2}\, dr={1\over w},
\end{equation}
that one bond out of $w$ has has radius $r_{\rm max}$ or larger.  This gives
$r_{\rm max}^{(b)}=s\sqrt{\ln w}$.  Similarly, the largest particle out of
$N$ has radius $r_{\rm max}^{(p)}=\sqrt{\ln N}$.  Thus $N_1\approx w^{s^2}$
particles typically need to be injected before one occurs which is large
enough to block the largest bond.  Continuing this picture sequentially, the
radius of the $k^{\rm th}$-largest bond is given by Eq.~(\ref{rmax}), but
with $1/w$ replaced by $k/w$; this gives $r_k= s\sqrt{\ln(w/k)}$.  Therefore
$N_k\approx (w/k)^{s^2}$ particles need to be injected before the $k^{\rm
  th}$-largest bond is blocked.  Since $N_k$ decreases rapidly with $k$, the
initial blockage events control the clogging time $T$, whose lower bound is
given by $T\approx N_1/w > w^{s^2-1}$.  The factor of $1/w$ arises because
the particle injection rate is proportional to $w$ for a fixed-density
suspension and a fixed pressure drop.  However, there is considerable particle
penetration in the large-pore limit, so that the equivalence to the clogging
of a parallel bond array is not really appropriate and the bound on $T$ is
quite crude.

The latter case where pores are smaller than particles ($s\alt 1$) is
simpler, as each particle injection event typically leads to the clogging of
the first pore entered.  Now the initial pores are blocked quickly, while
later pores are blocked more slowly because the overall flow rate decreases
significantly near the end of the clogging process.  We argue that clogging
is dominated by the time of these later blockage events.  First, let us
determine how the flow rate varies as the last few bonds get blocked.  Since
only the smallest bonds remain open near clogging, the permeability is
determined by these smallest radii.  We estimate the radius of the $k^{\rm
  th}$ smallest bond from
\begin{equation}
\label{kth}
\int_0^{r_k} 2\alpha r\, e^{-\alpha r^2}\,dr ={k\over w},
\end{equation}
which gives $r_k=s\sqrt{k/w}$.  The permeability of a parallel bundle of
these $k$ smallest pores is then
\begin{equation}
\label{K}
\kappa(k)=\sum_{j=1}^k r_j^4 \approx s^4\sum_{j=1}^k(j/w)^2 \sim s^4 k^3/w^2,
\end{equation}
while the initial system permeability (obtained by setting $k=w$ above) is
simply $\kappa(w)=s^4 w$.

Since the overall fluid flow is proportional to the permeability for a fixed
pressure drop, the time increment $t_k$ between blocking the $(k-1)^{\rm
  st}$-smallest and $k^{\rm th}$-smallest pore scales as
\begin{equation}
\label{tk}
t_k ={\kappa(w)\over w\kappa(k)}\sim {w^2\over k^3}.
\end{equation}
Here the factor $w$ in the denominator again accounts for a particle
injection rate proportional to $w$.  The clogging time $T$ is now dominated
by the time to block the smallest bond, so that $T > t_1 \propto w^2$.

Thus the relative pore and particle radii drives a transition in the clogging
process.  For $s^2>3$, clogging is dominated by the initial blockage events
and $T> w^{s^2-1}$, while for $s^2<3$, the last events dominate, leading to
$T> w^2$.  For the small-pore limit, we may carry the analysis further and
obtain the distribution of clogging times.  This distribution gives a {\em
  divergent\/} mean clogging time; nevertheless, suitably defined moments of
the clogging time distribution scale as $w^2$.

We find the clogging time distribution in terms of the radius distribution of
the smallest bond, since this bond ultimately controls clogging in a single
parallel bond array.  For the Hertz distribution, the probability that a
given bond has a radius greater than or equal to $r$, $B_>(r)$, is
\begin{equation}
\label{p>}
B_>(r)=\int_r^\infty 2\alpha r e^{-\alpha r^2}\,dr=e^{-\alpha r^2}.
\end{equation}
Then the radius distribution of the smallest bond from among $w$, $S_w(r)$, is
given by\cite{extreme}
\begin{equation}
\label{Sw}
S_w(r)= w\, b(r)\, B_>(r)^{w-1}= 2\alpha wr\, e^{-\alpha w r^2}.
\end{equation}
The first equality expresses the fact that one of the $w$ bonds has
(smallest) radius $r$, with probability $b(r)$, and the other $w-1$ must have
radii larger than $r$.

{}From the basic connections between permeability, pore radius, and time
scale (Eqs.~(\ref{K}) and (\ref{tk})), and the fact that the clogging time is
dominated by $t_1$, we deduce 
\begin{equation}
\label{connect}
T\sim t_1 \approx {\kappa(w)\over w\kappa(1)}= {s^4\over r_1^4},
\end{equation}
while the clogging time distribution, $P_w(T)$, is directly related
to the smallest bond radius distribution through $P_w(T)\,dT=S_w(r)\, dr$.
Using Eqs.~(\ref{Sw}) and (\ref{connect}), we obtain the basic result
(independent of system length)
\begin{equation}
\label{t-dist}
P_w(T)\sim {w\over T^{3/2}}\,\, e^{-w/T^{1/2}}.
\end{equation}
The power law should apply in the time range $w^2<t<w^2N^2$.  The former
corresponds to the size of the typical smallest pore in a single realization,
$s/\sqrt{w}$, while the latter corresponds to the smallest pore from among
$N$ realizations of the system, $s/\sqrt{Nw}$.  The essentially singular
short-time cutoff arises from those realizations where the smallest bond
happens to be anomalously large.  Other coincident particle and bond radius
distributions lead to qualitatively similar forms for $P_w(T)$, but with a
different exponent value.  For example, if $S_w(r)\sim r^\mu$ as $r\to 0$,
then the long-time exponent in $P_w(T)$ is $-(\mu+5)/4$. 

We test our predictions by Monte Carlo simulations of the motion and trapping
of suspended particles in a lattice porous medium.  Each bond (with unit
length) corresponds to a pore and the sites represent pore junctions.  The
algorithm is event-driven to bypass the physical slowing down of the flow as
clogging is approached.  At an intermediate stage, there are a finite number
of particles in the network, consistent with a unit pressure gradient and a
constant-density suspension.  Particles are defined to be always at lattice
sites and each (including newly injected particles) evolves by either: (a)
moving to the next ``downstream'' site, (b) blocking the downstream bond it
has entered, or (c) remaining stationary.  The probability of each of these
possibilities is defined so that the process corresponds to particles moving,
on average, at the local velocity.

To achieve this, each particle at a junction selects an outgoing bond $i$,
with entrance probability equal to the fractional flux into this bond.  In an
update, a particle attempts to move to the next junction, via the
pre-selected bond, with probability proportional to $v_i/v_{\rm max}$, where
$v_i$ is the local bond velocity and $v_{\rm max}$ is the instantaneous
largest particle velocity in the system.  If the attempt occurs, the particle
either moves to the next downstream site, or, if the particle is too large,
blocks the bond.  After a single update of all particles, the time is
incremented by $\Delta t=1/v_{\rm max}$ to ensure that, on average, a
particle moving along bond $i$ has speed $v_i$.

After all particles undergo their move attempts, ${\cal N}\propto \phi
\,\Delta t$ new particles are injected at the upstream end of the network,
where $\phi$ is the overall fluid flux.  This ensures the correct absolute
velocity for each particle.  Since the time increment systematically
increases as bonds get blocked, the particle injection rate progressively
slows.  Our approach is distinct from earlier multiparticle simulations of
filtration, where the injection rate was decoupled from the overall
flow\cite{sahimi}.

\begin{figure}
  \narrowtext \epsfxsize=2.2in\epsfysize=2.2in \hskip 0.3in\epsfbox{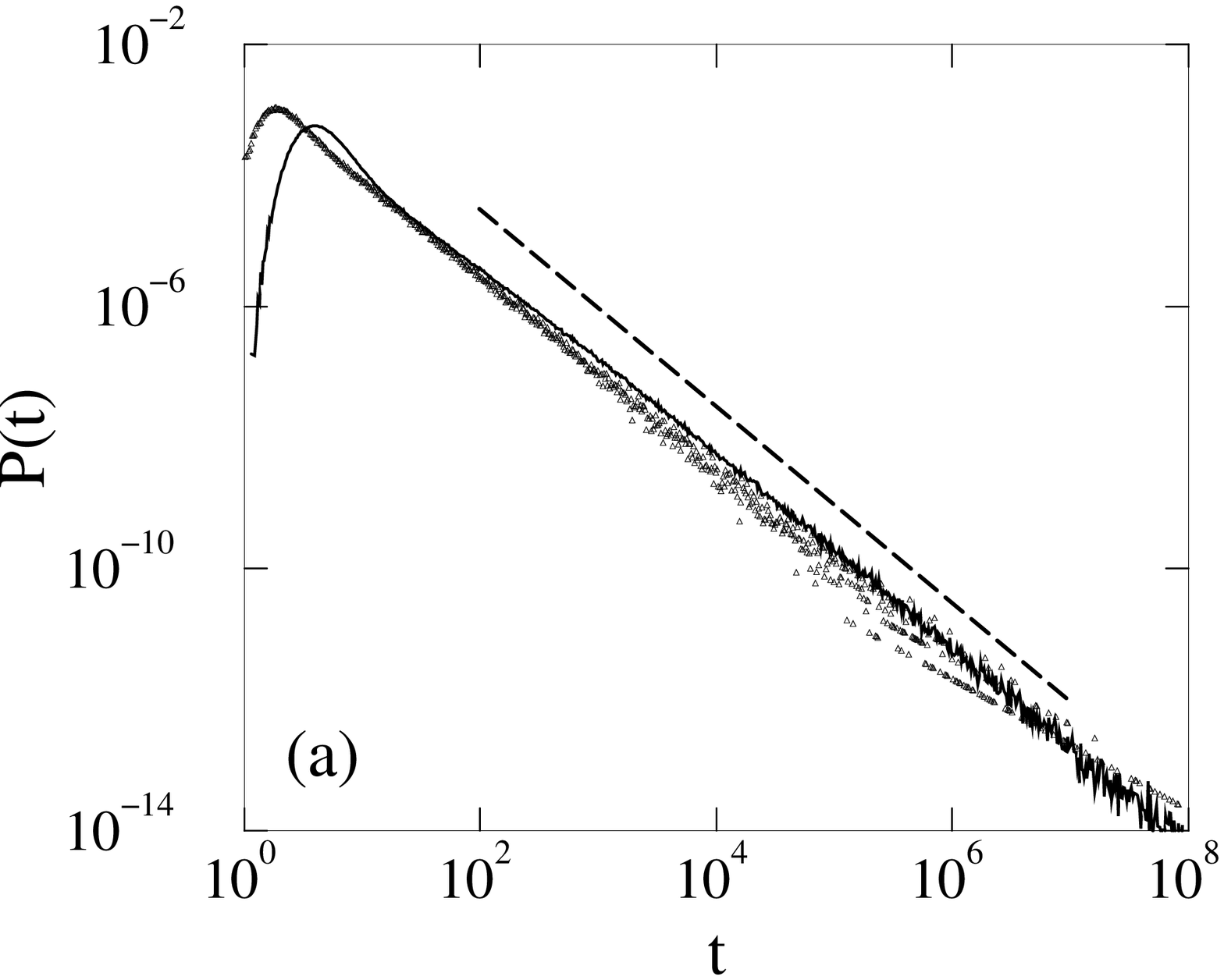}
  \narrowtext \epsfxsize=2.3in\epsfysize=2.3in \hskip 0.2in\epsfbox{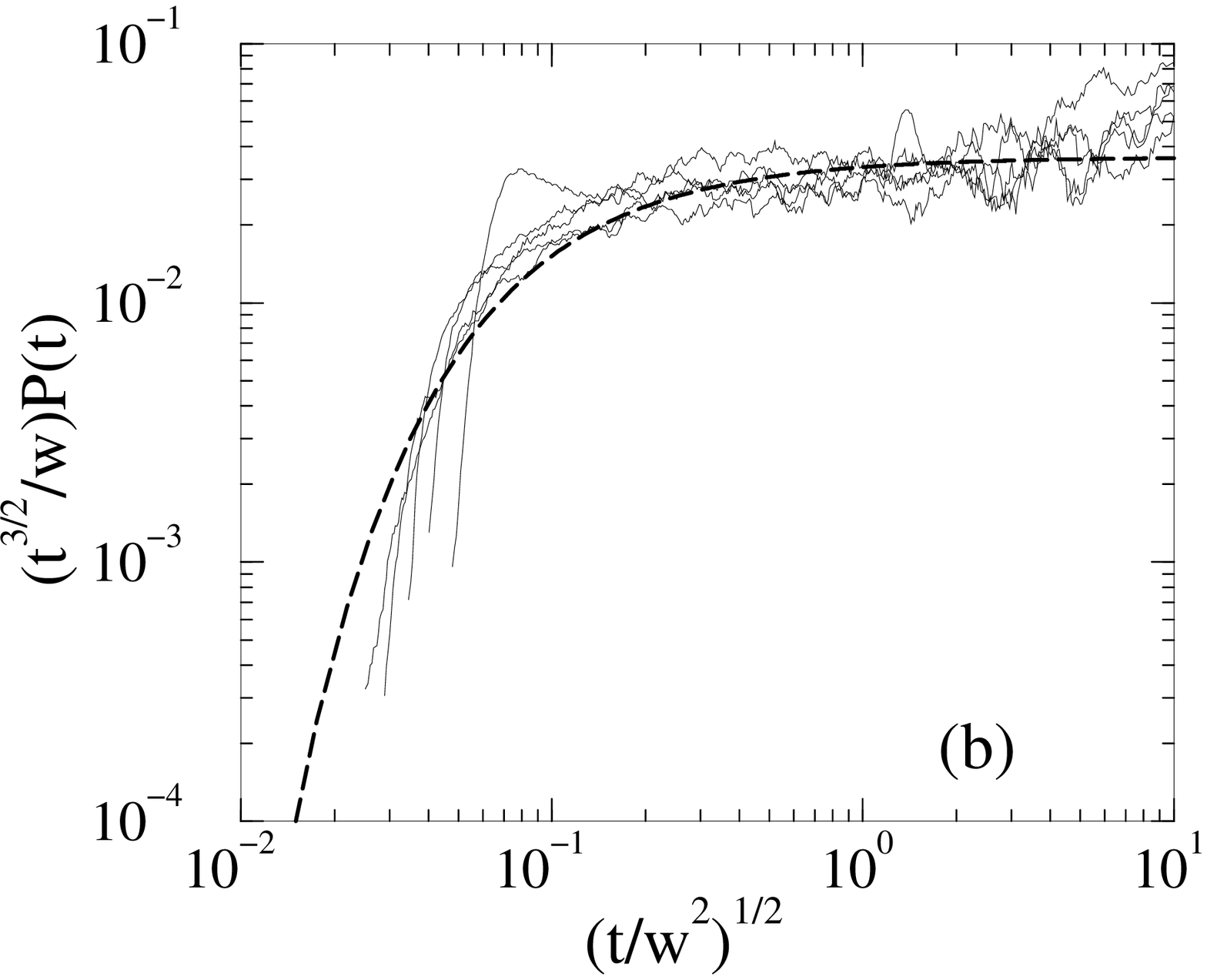}
  \vskip 0.15in
\caption{(a) Distribution of clogging times on the bubble model (continuous curve) 
  and on a square lattice (data points) for system width 20 and length 10.
  Time units correspond to an initial injection rate of 0.05 particles per
  unit width per unit time.  The straight line has slope $-3/2$.  (b) The
  scaled clogging time distribution for square lattices of length 10 and
  widths 20, 40, 60, 100, and 200 (right to left).  The data have been
  smoothed over a 12-point range.  The heavy dashed line is the prediction of
  Eq.~(\ref{t-dist}).
\label{Pt}}
\end{figure}

In Fig.~\ref{Pt}(a), we plot the clogging time distribution from simulations
on the square lattice and the bubble model of the same spatial extent.  The
latter is a series of parallel bond arrays with perfect mixing at each
junction; this idealized system accounts for geometrical aspects of clogging
\cite{dr}.  The agreement between the two distributions is remarkably good,
suggesting that the schematic evolution proposed in Fig.~\ref{cartoon} is
quantitatively correct.  The tails of the two distributions in
Fig.~\ref{t-dist}(a) are well-fit by a power law with exponent $-3/2$.  As a
further test of Eq.~(\ref{t-dist}), we plot, in scaled units, data for the
clogging time distribution for systems of various widths $w$
(Fig.~\ref{Pt}(b)) and compare to the functional form in Eq.~(\ref{t-dist}).

\begin{figure}
  \narrowtext \epsfxsize=2.2in\epsfysize=2.2in \hskip 0.3in\epsfbox{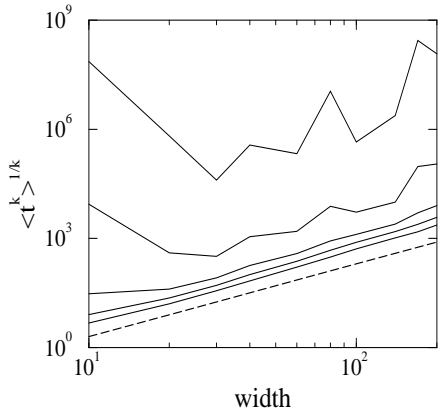}
  \vskip 0.15in
\caption{Width dependence of the moments $\langle
  t^k\rangle^{1/k}$ for $k=1, 1/2, 1/3, 1/4$, and 1/6 (top to bottom).  The
  dashed line corresponds to a $w^2$ dependence.
\label{moments}}
\end{figure}

An important consequence of the power-law tail in the clogging time
distribution is a transition in the corresponding moments.  These are
\begin{equation}
\label{t-int}
\langle t^k\rangle =\int_0^\infty t\, P_w(t)\, dt 
\approx \int_{w^2}^{N^2w^2} w\, t^{k-3/2}\, dt,
\end{equation}
where the approximation replaces the short-time cutoff at $t\approx w^2$ in
Eq.~(\ref{t-dist}) by the lower limit in the integral.  Consequently
\begin{equation}
\label{t-mom}
M_k(w)\equiv \langle t^k\rangle^{1/k}\sim \cases {w^2 N^{2-1/k} & $k>1/2$; \cr
              w^2 (\ln N)^2  & $k=1/2$; \cr w^2 & $k<1/2$.}
\end{equation}
Thus the mean clogging time diverges when the number of realizations is
infinite, leading to large sample to sample fluctuations in the clogging
time, while the moments $M_k$ with $k<1/2$ should be well behaved.  The
transitions in the moments is illustrated in Fig.~(\ref{moments}), where
$M_k$ behaves erratically for $k=1$ and 1/2, but then appears to grow as
$w^2$ for $k\leq 1/3$.

\begin{figure}
  \narrowtext \epsfxsize=2.2in\epsfysize=2.2in \hskip 0.3in\epsfbox{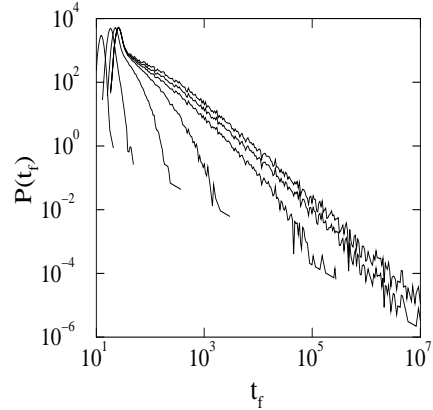}
  \vskip 0.15in
\caption{Probability that a filter with $w=200$ reaches permeability fraction 
  $f$ at time $t_f$, for $f=4^{-n}$, with $n=2, 4, 6, 8, 12, 16$, together
  with the time distribution until complete clogging (left to right).
\label{partial}}
\end{figure}

The behavior of $M_k$ raises the question of when is a filter no longer
useful.  Waiting until complete clogging is impractical because of large
fluctuations in the clogging time and eventual poor filter performance.  It
would be more useful to operate a filter only until the permeability decays
to a (small) fraction of its initial value, such that reasonable flow and
trapping efficiency are maintained, while minimizing fluctuations in this
threshold time.  We provide some preliminary results about these questions.
Simulations indicate that the time to reach permeability fraction $f$,
$t(f)$, is proportional to $w^{2/3}f^{-1/2}$ for $10^{-10}\alt f\alt
10^{-3}$, with the distribution of these fractional times progressively
broadening for decreasing $f$ (Fig.~\ref{partial}).  Thus waiting until the
permeability decays to a fixed fraction can provide at least a reliable
criterion for when a filter should be discarded.  A consequence of
$t(f)\propto w^{2/3}f^{-1/2}$ is that the permeability decays as $1/t^2$ over
a wide range.  Because of this rapid decrease, a practical filter should also
have pores typically larger than particles to have a reasonable lifetime.
These issues and understanding the efficiency of a filter throughout its
useful life are currently under investigation.

We thank P. L. Krapivsky for many helpful discussions and a critical
manuscript reading, as well as grants NSF DMR9978902 and ARO DAAD19-99-1-0173
for financial support.

\end{multicols} 
\end{document}